\documentclass[pra,12pt,tightenlines]{revtex4}

\usepackage{amsfonts}
\usepackage{graphicx}
\usepackage{bm}

\newcommand{\ket}[1]{|#1\rangle}
\newcommand{\bra}[1]{\langle#1|}
\newcommand{\braket}[2]{\langle#1|#2\rangle}
\newcommand{\ketbra}[2]{\ket{#1}\bra{#2}}
\newcommand{\tr}{\mbox{tr}}
\newcommand{\pc}{p^{\hspace{1pt}c}}
\newcommand{\pxj}{p^{\hspace{0.6pt}x,j}}
\newcommand{\pyj}{p^{\hspace{0.8pt}y,j}}
\newcommand{\Ec}{E^{\hspace{0.4pt}c}}
\newcommand{\Fc}{F^{\hspace{0.4pt}c}}
\newcommand{\I}{{\cal I}\hspace{0.3pt}}

\def\vec#1{\bm{#1}}

\begin{document}

\title{Minimal Informationally Complete Measurements for Pure States}
\author{Steven T. Flammia}
\author{Andrew Silberfarb}
\author{Carlton M. Caves}
\affiliation{Department of Physics and Astronomy, University of New
Mexico, Albuquerque, New Mexico 87131--1156, USA}
\date{April 22, 2004}

\begin{abstract}
We consider measurements, described by a positive-operator-valued
measure (POVM), whose outcome probabilities determine an arbitrary pure
state of a $D$-dimensional quantum system.  We call such a measurement
a {\it pure-state informationally complete\/} (PS\,I-complete) POVM. \
We show that a measurement with $2D-1$ outcomes cannot be
PS\,I-complete, and then we construct a POVM with $2D$ outcomes that
suffices, thus showing that a minimal PS\,I-complete POVM has $2D$
outcomes.  We also consider PS\,I-complete POVMs that have only
rank-one POVM elements and construct an example with $3D-2$ outcomes,
which is a generalization of the tetrahedral measurement for a qubit.
The question of the minimal number of elements in a rank-one
PS\,I-complete POVM is left open.
\end{abstract}

\maketitle

\section{Introduction}

An important technical requirement for developing coherent quantum
technologies is the ability to assess how well one can prepare or
create a particular quantum state.  This assessment is carried out by
making suitable measurements on a sequence of identically prepared
quantum systems.  The measurements are chosen so that the probabilities
of the measurement outcomes suffice to determine the state; the
probabilities are estimated from the outcome frequencies observed in
the measurements.  A set of measurements whose outcome probabilities
are sufficient to determine an arbitrary quantum state is called {\it
informationally complete\/} \cite{Prugovecki1977a,Busch1989a}.  The
process of determining the quantum state from the results of a sequence
of these measurements is called {\it quantum state tomography}.

Given a set of measurements that are informationally complete, they can
be amalgamated into a single generalized measurement, consisting of a
coin flip to select a measurement from the set, followed by an
application of that measurement. The statistics of such a generalized
measurement are described by a {\it positive-operator-valued measure\/}
(POVM) \cite{Peres1993a}, which consists of positive operators $\Ec$
that are labeled by an index $c$ that runs over the possible outcomes.
These operators satisfy a {\it completeness condition},
\begin{equation}
\sum_{c=1}^n\Ec=\mathbb{I}\;.
\label{eq:completeness}
\end{equation}
The operators $\Ec$, satisfying $0\le\Ec\le\mathbb{I}$, are called {\it
POVM elements.} The probability for outcome~$c$, given system state
$\rho$, is
\begin{equation}
\pc={\rm tr}(\rho \Ec)\;.
\label{eq:pc}
\end{equation}
The completeness condition and the unit-trace normalization of
$\rho$ guarantee that the probabilities are normalized to unity.

An informationally complete set of measurements can thus always be
reformulated as a single {\it informationally complete POVM\/}
(IC-POVM), i.e., a POVM having the property that the relations giving
$\pc$ in terms of $\rho$ can be inverted to give $\rho$ in terms of the
outcome probabilities.  A normalized density operator $\rho$ is
specified by $D^2-1$ real numbers, since an arbitrary Hermitian
operator on a $D$-dimensional quantum system is specified by $D^2$
independent real numbers, and this number is reduced by one for a
normalized density operator because it has unit trace.  Since the
relations~(\ref{eq:pc}) are linear, it is a simple matter of linear
algebra to conclude that an IC-POVM must provide $D^2-1$ independent
relations.  This might lead one to think that an IC-POVM can get by
with $D^2-1$ elements, but the normalization of the outcome
probabilities means that one of the relations~(\ref{eq:pc}) is
redundant, so an IC-POVM must have at least $D^2$ elements.

An equivalent, often more useful way of thinking is to disregard the
normalization of the density operator.  Then, like any Hermitian
operator, $\rho$ is specified by $D^2$ real numbers, the additional
number being ${\rm tr}(\rho)$.  By virtue of the completeness
condition, any POVM provides ${\rm tr}(\rho)$ as the sum of the
(unnormalized) probabilities~(\ref{eq:pc}).  This way of thinking
bypasses the waffling about normalization conditions and goes directly
to the point: an IC-POVM must provide $D^2$ linearly independent
relations and thus must have at least
$D^2$ POVM elements.

Informational completeness for arbitrary quantum states is thus
equivalent to the ability to reconstruct {\it any\/} Hermitian operator
$H$ from the operator inner products ${\rm tr}(H\Ec)$. This means that
the problem of finding informationally complete POVMs is reduced to
finding positive operators $\Fc$ that span the vector space of
Hermitian operators.  If these operators don't satisfy the completeness
condition~(\ref{eq:completeness}), one replaces them with the operators
$\Ec=G^{-1/2}\Fc G^{-1/2}$, where $G=\sum_c\Fc$ is trivially a
nonsingular operator. The new operators are positive, satisfy the
completeness condition, and are obviously informationally complete.
Using this procedure, it is easy to construct {\it minimal\/} IC-POVMs,
i.e., POVMs having $D^2$ elements, from sets of $D^2$ linearly
independent, positive operators. Indeed, it is easy to construct
minimal IC-POVMs for which all the POVM elements are rank-one, i.e.,
multiples of projectors onto pure states \cite{Caves2002b}.  Moreover,
there is evidence, both analytical and numerical, that in all finite
dimensions there are minimal, rank-one IC-POVMs, called {\it symmetric,
informationally complete POVMs\/} (SIC-POVMs), for which the operator
inner products of all pairs of POVM elements are the same
\cite{Renes2004a}.  Recast as ensembles of states, SIC-POVMs have
applications to quantum key distribution \cite{Renes2004b}, and they
have been shown \cite{Fuchs2004} to be the ensembles that are most
quantum according to a measure of quantumness introduced by Fuchs and
Sasaki \cite{Fuchs2003}.

In this paper we consider the pure-state version of informational
completeness, i.e., reconstruction of an arbitrary pure state
$\rho=|\psi\rangle\langle\psi|$ from the outcome probabilities
\begin{equation}
\pc=\langle\psi|\Ec|\psi\rangle\;.
\label{eq:pcpsi}
\end{equation}
We formalize the notion of pure-state information completeness in the
following definition.

\vspace{6pt}
\noindent
{\bf Definition (PS\,I-completeness).}  A {\it pure-state informationally
complete\/} ({\it PS\,I-complete\/}) POVM on a finite-dimensional quantum
system is a POVM whose outcome probabilities are sufficient to
determine any pure state (up to a global phase), except for a set of
pure states that is dense only on a set of measure zero.
\vspace{6pt}

\noindent
The intent of the last clause is to require the inversion to be unique
for generic pure states.  This clause says that any pure state outside
a set of measure zero is surrounded by an open ball of states all of which are
uniquely determined by the outcome probabilities.

Though some of the thinking and some of the mathematical techniques
developed for IC-POVMs can be transferred to a study of PS\,I-complete
POVMs, there is a critical difference, which makes PS\,I-completeness
more difficult to analyze: the relation between the outcome
probabilities and a density operator is linear, whereas the relation
between outcome probabilities and pure states is quadratic. The problem
of IC-POVMs thus lies squarely within linear algebra, whereas
PS\,I-completeness must be tackled using other tools.

The first point to make about PS\,I-completeness arises from a simple
counting argument.  A pure state is specified by $D$ complex
amplitudes, corresponding to $2D$ real numbers, but the number of
independent real numbers is reduced to $2D-2$ by the normalization
condition, $\langle\psi|\psi\rangle=1$, and by the fact that a pure
state can be multiplied by a global phase without changing any of its
predictions for probabilities.  Because of the redundancy due to the
normalization of the POVM probabilities, a PS\,I-complete POVM must
contain at least one more element than the $2D-2$ real numbers to be
determined.  This suggests that a PS\,I-complete POVM might get by with
only $2D-1$ elements.  This argument can be profitably rephrased in
terms of unnormalized pure states, which are specified by $2D-1$ real
numbers, giving immediately that a PS\,I-complete POVM must have at least
$2D-1$ elements.  Unlike the similar counting argument for density
operators, however, the counting argument for PS\,I-complete POVMs is
only suggestive, because linear algebra can't be applied to give
rigorous conclusions.  The counting argument leaves us only with a
prejudice that the minimal number of elements in a PS\,I-complete POVM
ought to be $2D-1$ or a little more.

The point of this paper is to provide some rigor.  We establish that
$2D$, not $2D-1$, is the minimal number of elements in a PS\,I-complete
POVM. \ To show this, we first prove that a POVM with fewer than $2D$
elements---in particular, one with $2D-1$ elements---cannot be
PS\,I-complete (Sec.~\ref{sec:necessity}), and we then construct an
example of a PS\,I-complete POVM with $2D$ elements
(Sec.~\ref{sec:sufficiency}).

Asher Peres considered a particular version of this problem in his
classic 1993 quantum-mechanics textbook \cite{Peres1993a}.  He noted
that the probabilities in two {\it complementary bases\/} should be
sufficient to determine a pure state up to a finite set of ambiguities.
If we let $\{|e_j\rangle,\;j=1,\ldots,D\}$, denote an orthonormal basis,
the complementary basis consists of the vectors
\begin{equation}
|f_k\rangle={1\over\sqrt D}\sum_{j=1}^D|e_j\rangle\,e^{2\pi ijk/D}\;.
\end{equation}
These complementary bases are discrete analogues of the position and
momentum bases of a particle moving in one spatial dimension.  A
consequence of our work is that measurements in two complementary bases
cannot be PS\,I-complete.  To see this, notice that the amalgamated POVM
for the two measurements has $2D$ POVM elements,
\begin{equation}
\Ec=
\cases{
{1\over2}|e_c\rangle\langle e_c|\;,&$c=1,\ldots,D$,\cr
{1\over2}|f_{c-D}\rangle\langle f_{c-D}|\;,&$c=D+1,\ldots,2D$.
}
\end{equation}
The factor of $1/2$ takes care of the coin flip that chooses with equal
probability between measurements in the two bases.  The probabilities
associated with the last POVM element in each basis,
$p^D={1\over2}|\langle e_D|\psi\rangle|^2$ and
$p^{2D}={1\over2}|\langle f_D|\psi\rangle|^2$, provide no information
because of the separate normalization of the measurements in the two
bases. Without losing any information, we can combine these two POVM
elements into a single element, ${1\over2}(|e_D\rangle\langle
e_D|+|f_D\rangle\langle f_D|)$, thus reducing the number of POVM
elements to $2D-1$.  Our proof in Sec.~\ref{sec:necessity} shows this
cannot be PS\,I-complete; attempts to find a pure state from the outcome
probabilities must at least suffer from the finite ambiguities
mentioned by Peres.  Following Peres, we illustrate these ambiguities
in two dimensions in Sec.~\ref{sec:twod}.

The problem considered by Peres is often called the {\it Pauli
problem}, as it is the finite-dimensional version of a question posed
by Pauli in a footnote to his quantum-mechanics article in {\it
Handbuch der Physik\/} \cite{Pauli}: can the wave function $\psi(x)$ of
a particle be determined (up to a global phase) from the position and
momentum probability distributions, $|\psi(x)|^2$ and
$|\tilde\psi(p)|^2$?  The answer to Pauli's question is a definitive
{\it no\/}: there are many wave functions not determined by their
position and momentum distributions and thus said to be {\it Pauli
nonunique}.   Much work has been devoted to investigating Pauli
nonunique states (see Refs.~\cite{Busch1989a,Weigert1992a,Weigert1996a}
and references cited therein), but even now there does not seem to be a
complete characterization of such wave functions.  A simple example of
a Pauli nonunique state is an eigenstate of parity,
$\psi(x)=\pm\psi(-x)$ (and $\tilde\psi(p)=\pm\tilde\psi(-p)$), that is
not time-reversal invariant, i.e., $\phi(x)=\psi^*(x)\ne
e^{i\alpha}\psi(x)$; such a wave function satisfies
$\tilde\psi^*(p)=\tilde\phi(-p)=\pm\tilde\phi(p)$, so $\psi(x)$ and
$\phi(x)$, though distinct states, have the same position and momentum
distributions. Discussion of work on Pauli nonunique states lies
outside the scope of our paper, but we do note the interesting result
of Corbett and Hurst \cite{Corbett1978a} that Pauli nonunique wave
functions are dense in the space of wave functions, a result that
motivates the denseness restriction in our definition of
PS\,I-completeness.

Previous work on PS\,I-completeness for finite-dimensional quantum
systems \cite{Weigert1992a,Amiet1999b,Amiet1999c} has considered the
system to be a spin-$s$ particle ($D=2s+1$) and has focused mainly on
reconstructing a pure state from the probabilities for spin components
along several directions.  The angular-momentum algebra and the
rotation group play essential roles in these considerations.  We review
some of these results during the course of our discussion and point out
how they are related to our work and, in particular, how they can be
rephrased in terms of a single POVM. \

Our paper is organized as follows.  In Sec.~\ref{sec:twod}, we
introduce informationally complete and PS\,I-complete POVMs in two
dimensions and note that the two-dimensional case provides little
guidance for generalizing PS\,I-completeness to higher dimensions.  In
Sec.~\ref{sec:necessity}, we prove that the outcome probabilites of a
POVM with $2D-1$ elements are insufficient to determine a general pure
state in $D$ dimensions, leading to, at best, a two-state ambiguity. In
Sec.~\ref{sec:sufficiency}, we construct a PS\,I-complete POVM with
$2D$ elements, and in Sec.~\ref{sec:pure}, we turn to the question of
rank-one PS\,I-complete POVMs and construct one with $3D-2$ POVM
elements.  Section~\ref{sec:conclusion} gives a brief summary of open
questions, while Sec.~\ref{S:epilogue} closes with a poetic description
of one of these questions, dedicated to Asher Peres on the occasion of
his 70th birthday\footnote{On a first reading, the reader can omit
Sec.~\ref{S:epilogue}; alternatively, the reader might prefer to skip
directly to Sec.~\ref{S:epilogue} and omit the rest of the article.}.

\section{Informationally complete POVMs in two dimensions}
\label{sec:twod}

We can gain some insight into IC-POVMs and PS\,I-complete POVMs by
looking at the case of a two-dimensional quantum system (qubit), where
the Bloch representation of states and operators permits us give
a complete characterization of information completeness.

The Bloch representation of an arbitrary POVM element for a qubit is
\begin{equation}
E=a\mathbb{I}+b\vec n\cdot\vec\sigma\;,
\end{equation}
where $\vec n$ is a unit vector in $\mathbb{R}^3$ and $a$ and $b$ are
nonnegative real numbers satisfying $b\le a$ and $b\le1-a$, to ensure
that $E$ and $\mathbb{I}-E$ are positive operators.  Rank-one POVM
elements have $a=b$.  A POVM is made up of $n$ such elements,
\begin{equation}
\Ec=a^c\,\mathbb{I}+b^c\vec n^c\cdot\vec\sigma\;,
\quad c=1,\ldots,n,
\end{equation}
satisfying the completeness condition~(\ref{eq:completeness}), which
now becomes two conditions,
\begin{eqnarray}
1&=&\sum_ca^c\;,\\
0&=&\sum_cb^c\vec n^c\;.
\label{eq:betan}
\end{eqnarray}

A POVM is informationally complete if and only if the POVM elements
$\Ec$ span the four-dimensional space of Hermitian operators.
Translated to the Bloch representation, this says that a POVM is
informationally complete if and only if the vectors $b^c\vec n^c$ span
$\mathbb{R}^3$.  A direct way to see this in the Bloch representation
is to note that for an arbitrary density operator
$\rho={1\over2}(\mathbb{I}+\vec P\cdot \vec\sigma)$, with polarization
vector $\vec P$ ($|\vec P|\le1$), the outcome probabilities are given
by
\begin{equation}
\pc=\tr(\rho\Ec)=a^c+b^c\vec n^c\cdot\vec P\;;
\label{eq:2dprobs}
\end{equation}
to reconstruct an arbitrary polarization vector, the vectors
$b^c\vec n^c$ must span $\mathbb{R}^3$.

The minimal number of elements in an IC-POVM is four; one sees this
directly in the Bloch representation by noting that the vectors
$b^c\vec n^c$ must span $\mathbb{R}^3$ and also satisfy the
condition~(\ref{eq:betan}), implying that there must be at least four
vectors.  For a minimal IC-POVM, it is also easy to see from
condition~(\ref{eq:betan}) that any three of the four vectors
$b^c\vec n^c$ must be linearly independent and thus span
$\mathbb{R}^3$ (and this means that none of the $b^c$ can be zero).

An example of a minimal, rank-one IC-POVM, which we make use of in
Sec.~\ref{sec:pure}, is given by the {\it tetrahedral measurement},
which is specified by the four unit vectors
\begin{eqnarray}
\vec n^1&=&\vphantom{{2\sqrt2\over3}}(0,0,1)\;,\nonumber\\
\vec n^2&=&\Biggl({2\sqrt2\over3},0,-{1\over3}\Biggr)\;,\nonumber\\
\vec n^3&=&\Biggl(-{\sqrt2\over3},\sqrt{{2\over3}},-{1\over3}\Biggr)\;,\nonumber\\
\vec n^4&=&\Biggl(-{\sqrt2\over3},-\sqrt{{2\over3}},-{1\over3}\Biggr)\;,
\label{eq:tetrahedron}
\end{eqnarray}
with $a^c=b^c=1/4$, $c=1,2,3,4$.  This POVM is also a SIC-POVM
by virtue of the symmetric placement of the Bloch vectors, which
connect the origin to the vertices of a tetrahedron whose apex is at
the north pole of the Bloch sphere.

A PS\,I-complete measurement is one such that the outcome
probabilities~(\ref{eq:2dprobs}) determine uniquely a generic pure
state $\ketbra{\vec m}{\vec m}={1\over2}(\mathbb{I}+\vec
m\cdot\vec\sigma)$.  It is immediately clear that to reconstruct an
arbitrary unit vector $\vec m$, the vectors $b^c\vec n^c$ must span
$\mathbb{R}^3$, so a PS\,I-complete measurement in two dimensions is
always informationally complete for all states, pure or mixed.  For
this reason PS\,I-completeness for qubits provides little guidance for
what happens in higher dimensions: in two dimensions, the $2D$ POVM
elements required for PS\,I-completeness provide the $D^2$ elements
necessary for informational completeness for all states, whereas in
higher dimensions, there is a yawning gap between $2D$ and $D^2$.

It is worth stressing the way a three-element POVM fails to be
PS\,I-complete. The condition~(\ref{eq:betan}) implies that the three
vectors $b^c\vec n^c$ are linearly dependent and thus span at most a
plane.  The outcome probabilities~(\ref{eq:2dprobs}) provide no
information about the component of $\vec m$ orthogonal to the plane.
When the vectors $b^c\vec n^c$ do span a plane, the outcome
probabilities determine the projection of $\vec m$ onto this plane,
which specifies everything about $\vec m$ except the sign of the
component orthogonal to the plane.  This two-fold ambiguity is the
source of the POVM's incompleteness for pure states.

A symmetric example of a three-outcome, rank-one measurement is the
{\it trine measurement\/} on a qubit \cite{Peres1992a}, given by the
unit vectors
\begin{eqnarray}
    \vec{n}^1 & = & \vphantom{{1\over2}}(1,0,0)\;, \nonumber\\
    \vec{n}^2 & = & \Biggl(-\frac{1}{2},\frac{\sqrt{3}}{2},0\Biggr)\;, \nonumber \\
    \vec{n}^3 & = & \Biggl(-\frac{1}{2},-\frac{\sqrt{3}}{2},0\Biggr)\;,
\end{eqnarray}
with $a^c=b^c=1/3$, $c=1,2,3$.  These vectors point from the origin to
the vertices of an equilateral triangle lying in the equatorial plane
of the Bloch sphere.  The outcome probabilities for the trine determine
the equatorial component of the Bloch vector for a pure state, but
can't resolve whether the Bloch vector is in the northern hemisphere or
the southern hemisphere, thus leaving a two-state ambiguity.

\section{Necessity of $2D$ or more outcomes}
\label{sec:necessity}

In this section we show that the outcome probabilities of a POVM with
$2D-1$ outcomes are insufficient to determine a generic pure state.  In
particular, we prove the following theorem.

\vspace{6pt}
\noindent
{\bf Theorem.}  The outcome probabilities
$\pc=\langle\psi|\Ec|\psi\rangle$ of a POVM with $2D-1$ measurement
outcomes are insufficient to determine a generic pure state
$|\psi\rangle$ in $D$ dimensions.

\vspace{6pt}
\noindent
We prove the theorem by showing that any allowed distribution of
outcome probabilities, except a set of distributions of measure zero,
is consistent with at least two possible pure states.  The proof
occupies the remainder of this section.

We begin by picking a particular POVM element, say $E^1$, and writing
it in its orthonormal eigenbasis $\{|j\rangle,\;j=0,\ldots,D-1\}$:
\begin{equation}
E^1=\sum_{j=0}^{D-1}\lambda_j|j\rangle\;.
\end{equation}
We order the eigenvalues from smallest to largest, i.e.,
$0\le\lambda_0\le\lambda_1\le\ldots\le\lambda_{D-1}$.  We can assume
that $\lambda_{D-1}>\lambda_0$, because otherwise $E^1$ would be a
multiple of the identity operator and would provide no useful
information beyond that always contained in the normalization
constraint.  Any pure state can be expanded in this eigenbasis as
\begin{equation}\label{E:pure}
     \ket{\psi} = r_0 \ket{0} + \sum_{j=1}^{D-1} c_j \ket{j} \;.
\end{equation}
We use the global phase freedom to make $r_0$ real, but we do not
require $\ket{\psi}$ to be normalized.  Instead we consider the norm to
be an extra parameter of the state, which is determined by the sum of
the (unnormalized) outcome probabilities~(\ref{eq:pcpsi}),
\begin{equation}
\sum_{c=1}^{2D-1}\pc=\braket{\psi}{\psi}\equiv R^2\;,
\label{eq:R}
\end{equation}
by virtue of the completeness condition~(\ref{eq:completeness}) for our
POVM. \

The amplitudes $c_j$ are arbitrary complex numbers, which we now write
in terms of their real and imaginary parts, i.e., $c_j= x_j+iy_j$.  The
variables necessary to define the state are now the $2D-1$ real
co-ordinates $r_0$, $x_1,\ldots,x_{2D-1},y_1,\ldots,y_{2D-1}$,
each of which is free to take on any real value.  We find it convenient
to put all these co-ordinates into a single real vector
$\vec\xi\in\mathbb{R}^{2D-1}$, whose components are
\begin{eqnarray}
    \xi_0&=&r_0 \;, \nonumber\\
    \xi_j&=&x_j\quad\text{and}\quad\xi_{j+D-1}=y_j\;,\quad j=1,\ldots,D-1.
\end{eqnarray}
Notice that an inversion through the origin, i.e.,
$\vec\xi\rightarrow-\vec\xi$, produces a global sign change in
$|\psi\rangle$.  When $r_0=\xi_0\ne0$, we could remove the global phase
freedom entirely by making $\xi_0$ positive, but we choose to retain
this two-fold ambiguity so that the space we are dealing with is
the entirety of $\mathbb{R}^{2D-1}$.

Each of the $2D-1$ outcome probabilities can now be written as a
quadratic form
\begin{equation}
    \pc= \bra{\psi}\Ec\ket{\psi} = \sum_{r,s}\xi_r M^c_{rs}\, \xi_s\;,
    \label{eq:drew}
\end{equation}
where the quantities $M^c_{rs}$ are the matrix elements of a real,
symmetric, positive (semi-definite) matrix ${\sf M}^c$.  A surface of
constant $\pc>0$ is a $2(D-1)$-dimensional (possibly degenerate)
hyperellipsoid ${\cal E}_{\pc}$, which extends to $\pm\infty$ in a
number of dimensions given by $2D-1-{\rm rank}({\sf M}^c)$ and has a
hyperellipsoidal cross-section in the remaining dimensions, numbering
${\rm rank}({\sf M}^c)$.  Notice that by construction the first outcome
probability has a diagonal matrix ${\sf M}^1$,
\begin{equation}
p^1=\lambda_0 r_0^2+\sum_{j=1}^{D-1}\lambda_j|c_j|^2
=\lambda_0\xi_0^2+\sum_{j=1}^{D-1}\lambda_j(\xi_j^2+\xi_{j+D-1}^2)\;,
\label{eq:p1}
\end{equation}
and is thus a hyperellipsoid aligned with the co-ordinate axes.

We now replace two of the outcome probabilities by equivalent
constraints.  First we replace the last
outcome probability ($c=2D-1$) by the normalization condition~(\ref{eq:R}),
\begin{equation}
R^2=\langle\psi|\psi\rangle=\sum_{r=0}^{2D-2}\xi_r^2\;,
\end{equation}
something we can do because
$p^{2D-1}=R^2-\sum_{c=1}^{2D-2}\pc$. A surface of constant
norm $R^2$ is, of course, a sphere ${\cal S}_R$ of radius $R$ in
$\mathbb{R}^{2D-1}$.  Each physical state $|\psi\rangle$ with
$\xi_0\ne0$ is represented twice on this sphere of radius $R$, once in
the northern hemisphere ($\xi_0>0$) and again, with its sign reversed,
at the antipode in the southern hemisphere ($\xi_0<0$).  States on the
equator ($\xi_0=0$) retain the full global phase freedom.

The second modification is to replace the first outcome
probability~(\ref{eq:p1}) with an equivalent constraint in which
$\xi_0$ does not appear:
\begin{equation}
p^1-\lambda_0R^2
=\sum_{j=1}^{D-1}(\lambda_j-\lambda_0)(\xi_j^2+\xi_{j+D-1}^2)
\equiv P\ge0\;.
\label{eq:P}
\end{equation}
That $P\ge0$ follows because $\lambda_0$ is the smallest eigenvalue.
Moreover, we are guaranteed that $P$ is not always zero because
$\lambda_{D-1}>\lambda_0$.  All this means that a surface of constant
$P>0$ is a hyperellipsoid ${\cal E}_P$, aligned with the $\xi_r$ axes,
which runs off to $\pm\infty$ along the $\xi_0$ axis (and perhaps other
axes) and which has an ellipsoidal boundary in one or more co-ordinates
$\xi_r$.

Now consider a particular set of values for $R^2$,
$p^{2D-2},\ldots,p^2$, and $P$, and imagine applying the constraints
imposed by these $2D-1$ measured values in the order listed.  The first
step in the process yields the submanifold ${\cal I}_1\equiv {\cal
E}_{R^2} = {\cal S}_R $.  The set $\I_m$ obtained after $m$ steps is
the intersection of $\I_{m-1}$ with the hyperellipsoidal submanifold
${\cal E}_{p^{2D-m}}$,
\[ \I_m = \I_{m-1} \cap {\cal E}_{p^{2D-m}} \; . \]  The last step
intersects $\I_{2D-2}$ with ${\cal E}_P$ to yield the set
$\I_{2D-1}=\I_{2D-2}\cap{\cal E}_P$.  We note that for all $m$, $\I_m$
is a compact set because the process starts with the sphere ${\cal
S}_R$, which is itself compact.  For PS\,I-completeness the final set
$\I_{2D-1}$ must consist generically of just the two antipodal points
representing a particular pure state.

\begin{figure}
\includegraphics{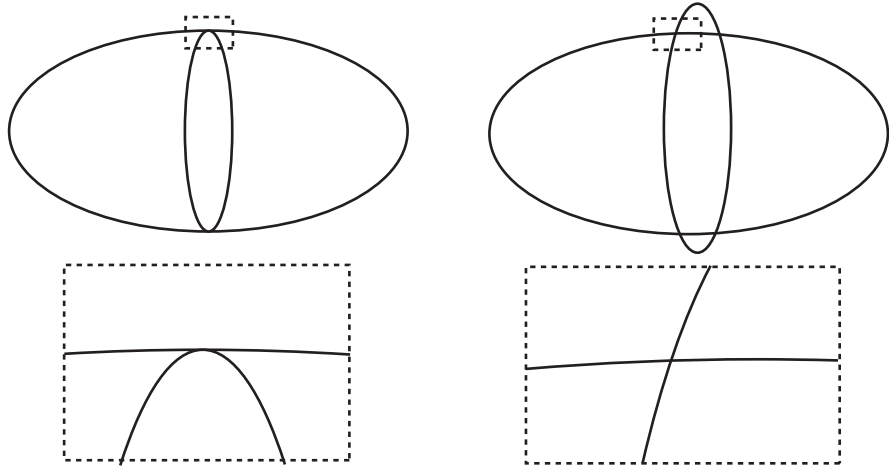}
\caption{\label{F:transversal} Transversal and nontransversal
intersections.  The two ellipses shown in each subfigure are
submanifolds of $\mathbb{R}^2$.  {\it a)\/}~shows a nontransversal
intersection of two ellipses, and {\it b)\/} is a close-up of one of
the intersection points.  The intersection in {\it b)\/} is {\it
non\/}transversal because the tangent vectors at the point shown are
the same; they therefore span only a one-dimensional subspace of
$\mathbb{R}^2$.  {\it c)\/} shows a slight perturbation of {\it a)\/},
and by looking at the close-up of the intersection in {\it d)\/}, we
see that the intersection is now transversal.  This is because at all
points of intersection, the tangent vectors of the two submanifolds
span $\mathbb{R}^2$.}
\end{figure}

Having set up the entire problem, we now need a bit of differential
geometry about the intersection of submanifolds.  Given a manifold $R$
of dimension $d_R$, two submanifolds, $M$ and $N$, of dimensions $d_M$
and $d_N$, intersect {\it transversally\/} if at each point of
intersection ${\cal P}$, the vectors in the tangent spaces to $M$ and
$N$, $T_M({\cal P})$ and $T_N({\cal P})$, span the entire tangent space
to $R$ at ${\cal P}$ (see Fig.~\ref{F:transversal}).  A transversal
intersection $M\cap N$ is a submanifold of dimension $d_{M\cap
N}=d_M+d_N-d_R$, since this relation holds for the dimensions of the
tangent spaces at each point of intersection.  A fundamental theorem of
differential geometry, called {\it Sard's theorem\/} \cite{sard},
asserts that if two submanifolds with a nonempty intersection intersect
nontransversally, they can be perturbed slightly to intersect
transversally. Figure~\ref{F:transversal} illustrates the difference
between transversal and nontransversal intersections and how a
perturbation of a nontransversal intersection yields a transversal
intersection.

We now apply these ideas to our situation.  Notice that if the
successive intersections in our sequence are transversal, then each
intersection produces a new submanifold with dimension decreased by
one, $d_{\I_m}=d_{\I_{m-1}}+(2D-2)-(2D-1)=d_{\I_{m-1}}-1$, so the
dimension of $I_m$ is $2D-1-m$.  We now argue that if the measurement
is to be PS\,I-complete, a sequence of such transversal intersections
must occur generically, i.e., for all outcome probabilities $p^c$
except for a set of measure zero.

\begin{figure}
\includegraphics{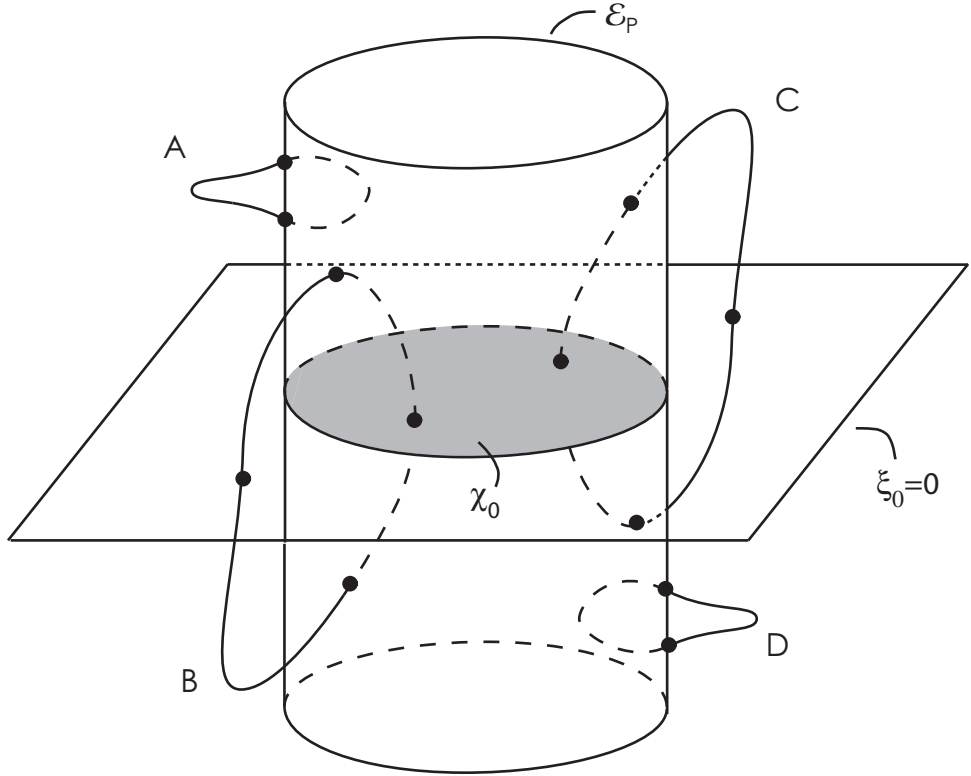}
\caption{\label{F:solutions} The cylinder is the hyperellipsoidal
submanifold ${\cal E}_P$ with many dimensions suppressed.  Recall that
${\cal E}_P$ extends to infinity, though it is drawn as finite here.
The union of the closed curves $A$, $B$, $C$, and $D$ is the
submanifold $\I_{2D-2}$.  The convention for dashed and dotted lines is
that lines that are behind one layer of the cylinder are dashed, and
lines that are behind two layers are dotted.  The points where
$\I_{2D-2}$ intersects ${\cal E}_P$ (denoted as black dots) are the
(zero-dimensional) submanifold $\I_{2D-1}$.  Notice that the entire
diagram is invariant under inversion through the origin. The shaded
disk ${\cal X}_0$, which is the part of the $\xi_0=0$ hyperplane that
is in or on the interior of the ${\cal E}_P$ submanifold, also
has inversion-symmetric points of intersection with $\I_{2D-2}$.  The
part of ${\cal E}_P$ with $\xi_0>0$, denoted in the text as ${\cal
E}^{(+)}_P$, has multiple intersections with $\I_{2D-2}$, implying that
Eqs.~(\ref{eq:drew}) cannot be inverted to reconstruct a unique pure
state $\ket{\psi}$.}
\end{figure}

Suppose instead that the intersections are transversal up to the $m$th
step, at which step $\I_{m-1}$ and ${\cal E}_{p^{2D-m}}$ intersect
nontransversally.  There are two ways this could happen. First, it
could be that part or all of $\I_{m-1}$ lies in ${\cal E}_{p^{2D-m}}$,
so that part or all of the intersection fails to have dimension reduced
by one.  If this situation were generic, it would mean that the final
intersection would be greater than zero-dimensional, thus not
specifying points, so the POVM could not be PS\,I-complete.  Second, it
could be that part or all of the intersection has dimensionality
reduced by more than one, as in the case of two ellipsoids touching at
tangent points.  This is the situation depicted in Figure
\ref{F:transversal}{\it a)\/} if the ellipses were replaced with
surfaces of revolution about the vertical axis in the diagram.  If this
situation were generic, it would imply a relation between the $m$
remaining values, $p^{2D-m},\ldots,P$, and the previous values,
$R^2,\ldots,p^{2D-m+1}$; since an arbitrary state is not subject to
such a restriction, the POVM could not be PS\,I-complete.  We conclude
that if a measurement is to be PS\,I-complete, a transversal intersection
at every step in our process is generically necessary.  In accordance
with Sard's theorem, a nontransversal intersection at any step could be
removed by slightly perturbing the values of our constraints.

Now consider the final step in our process.  As we have just argued, if
the POVM is to be PS\,I-complete, this step generically involves the
transversal intersection of a compact, one-dimensional submanifold,
$\I_{2D-2}$, with the hyperellipsoid ${\cal E}_P$, which is
defined by Eq.~(\ref{eq:P}) (with $P>0$ generically), and the result of
this final intersection is a set of points, $\I_{2D-1}$.  The
submanifold $\I_{2D-2}$, being compact and one-dimensional, is a
set of closed curves that do not intersect one another or themselves.
Moreover, the hyperellipsoid ${\cal E}_P$ is closed and orientable,
thus splitting $\mathbb{R}^{2D-1}$ into an inside and an outside.  A
closed curve intersecting such a surface transversally must intersect
it an even number of times, since any entrance from the outside must be
paired with an exit.  Thus we can conclude that $\I_{2D-1}$
consists generically of an even number of points.

We need, of course, a stronger result than this because the inversion
symmetry, $\vec\xi\rightarrow-\vec\xi$, already implies that the points
in $\I_{2D-1}$ come in pairs.  What we need to show is that $\I_{2D-1}$
contains at least two points with $\xi_0>0$, but we can get this by a
simple extension of the above reasoning, which is illustrated in
Fig.~\ref{F:solutions}.  Let ${\cal X}_0$ be the part of the hyperplane
$\xi_0=0$ that lies inside or on ${\cal E}_P$.  The inversion symmetry
guarantees that the points in the intersection of $\I_{2D-2}$ with
${\cal X}_0$ come in pairs, $\vec\xi$ and $-\vec\xi$. The points in a
pair are distinct, because an intersection at $\vec\xi=0$ is forbidden
by the normalization constraint.

Now consider the union of the $\xi_0>0$ part of ${\cal E}_P$, which we
denote as ${\cal E}_P^{(+)}$, and ${\cal X}_0$.  Since this union is
closed and orientable, we can conclude, just as we did above for ${\cal
E}_P$, that the closed curves in $\I_{2D-2}$ intersect it an even
number of times.  Since we have just established that $\I_{2D-2}$
intersects ${\cal X}_0$ an even number of times, we find that
$\I_{2D-2}$ intersects ${\cal E}_P^{(+)}$ an even number of times. This
means that the final set, $\I_{2D-1}$, generically has an even number
of points with $\xi_0>0$, corresponding to distinct pure states
consistent with the outcome probabilities.  This completes the proof of
Theorem~1, because this generic ambiguity, at least two-fold, shows
that a POVM with $2D-1$ elements cannot be PS\,I-complete.

\section{Sufficiency of $2D$ outcomes}
\label{sec:sufficiency}

In this section we construct an explicit example of a PS\,I-complete POVM
with $2D$ outcomes, thus showing the sufficiency of this number of
outcomes for PS\,I-completeness.  Combined with the result of the
preceding section, this shows that a {\it minimal\/} PS\,I-complete POVM
has $2D$ elements.

Letting $\{|j\rangle,\;j=0,\ldots,D-1\}$ denote an orthonormal basis
for a $D$-dimensional quantum system, we define the operators
\begin{eqnarray}
X_{jk} &\equiv& \ketbra{j}{k} + \ketbra{k}{j}\;,\nonumber\\
Y_{jk} &\equiv& -i\ketbra{j}{k}+ i\ketbra{k}{j}\;,
\label{eq:Paulijk}\\
Z_{jk} &\equiv& \ketbra{j}{j} - \ketbra{k}{k}\;.\nonumber
\end{eqnarray}
These operators can be thought of as the Pauli operators in the
subspace spanned by $\ket{j}$ and $\ket{k}$.  In this section we only
need the $0j$ versions of these operators, but we use others in
Sec.~\ref{sec:pure}.

We now consider a POVM consisting of the following positive operators:
\begin{eqnarray}
&\mbox{}&a\ketbra{0}{0}\;,\nonumber\\
&\mbox{}&b(\mathbb{I}+X_{0j})\quad\text{and}\quad b(\mathbb{I}+Y_{0j})\;,
\quad\mbox{$j=1,\ldots,D-1$,}
\label{eq:2DPOVM}\\
&\mbox{}&T\;.\nonumber
\end{eqnarray}
Here $a$ and $b$ are positive numbers, and $T$ is a ``throw-away'' POVM
element that must be included to satisfy the completeness
condition~(\ref{eq:completeness}).  We can always make $T$ a positive
operator by choosing $a$ and $b$ small enough. Notice that this POVM
has $2D$ elements, as promised.  We now demonstrate that this POVM is
PS\,I-complete.

We let
\begin{equation}
\ket{\psi} = \sum_{j=0}^{D-1} c_j \ket{j}
\label{eq:psi}
\end{equation}
be an arbitrary normalized pure state, expanded in the basis
$\{\ket{j}\}$.  We remove the global phase freedom by choosing
$c_0=r_0$ to be real and nonnegative, and we write $c_j=x_j+iy_j$ for
the remaining amplitudes.

The first outcome probability for the POVM~(\ref{eq:2DPOVM}) is
\begin{equation}
p^0=a|\braket{0}{\psi}|^2=a r_0^2 \;,
\end{equation}
from which we get the (positive) amplitude of the state $\ket{0}$,
\begin{equation}
r_0=\sqrt{\frac{p^0}{a}}\;.
\end{equation}
The remaining outcome probabilities (except for the throw-away outcome)
are
\begin{eqnarray}
    \pxj&=&b\bra{\psi}(\mathbb{I}+X_{0j})\ket{\psi}=b(1+2r_0x_j)\;, \nonumber\\
    \pyj&=&b\bra{\psi}(\mathbb{I}+Y_{0j})\ket{\psi}=b(1+2r_0y_j)\;,
\label{eq:pxy}
\end{eqnarray}
for $j=1,\ldots,D-1$.  Except for states with $r_0=0$ (a set of measure
zero), these, too, can be immediately inverted to give the remaining
amplitudes,
\begin{eqnarray}
x_j&=&\frac{\pxj-b}{2b r_0}\;,\\
y_j&=&\frac{\pyj-b}{2 b r_0}\;.
\end{eqnarray}
This completes the (trivial) demonstration that the
POVM~(\ref{eq:2DPOVM}) is PS\,I-complete.

The inversion procedure fails on the $(D-1)$-dimensional subspace
orthogonal to $\ket{0}$, where $r_0=0$.  There is a simple way to
handle this failure in a tomographic procedure, i.e., in a sequence of
measurements of this POVM, made on many identical copies of the same
system, with the outcome probabilities estimated from the outcome
frequencies. All one has to do is to precede the tomographic procedure
with a single premeasurement in the basis $\{|j\rangle\}$.  Whatever
the outcome of this premeasurement, that outcome has nonzero
probability, so we can choose it to be the state~$|0\rangle$.

Notice that after determining $r_0$, the reconstruction of the
remaining amplitudes $c_j$ is reduced to solving linear equations. This
transformation of a fundamentally quadratic inversion into a linear one
is the reason it is so easy to show that the POVM~(\ref{eq:2DPOVM}) is
PS\,I-complete. The key to this transformation is that the
probabilities~(\ref{eq:pxy}) have no terms that are quadratic in the
amplitudes $c_j$.  This, in turn, is a consequence of having the unit
operator in the ``middle'' POVM elements, $b(\mathbb{I}+X_{0j})$ and
$b(\mathbb{I}+Y_{0j})$, since for normalized states, these unit
operators always put a 1 into the corresponding outcome probabilities.
We can't just leave the unit operator out, because we have to put
something there to make the middle POVM elements positive.  Suppose, for
example, that we tried to make the middle POVM elements rank one by
changing them to $b(P_{0j}+X_{0j})$ and $b(P_{0j}+Y_{0j})$, where
$P_{0j}=\ketbra{0}{0}+\ketbra{j}{j}$ is the projection operator onto
the subspace spanned by $\ket{0}$ and $\ket{j}$.  Then the
probabilities~(\ref{eq:pxy}) become $\pxj=b(r_0^2+x_j^2+y_j^2+2r_0x_j)$
and $\pyj=b(r_0^2+x_j^2+y_j^2+2r_0y_j)$.  The quadratic term
$x_j^2+y_j^2$ introduces a generic two-fold ambiguity into the
determination of each of the amplitudes $c_j$, which makes the POVM not
PS\,I-complete.

We should stress that the role of the POVM~(\ref{eq:2DPOVM}) is solely
to demonstrate that $2D$ outcomes are sufficient for PS\,I-completeness,
because the presence of the unit operator in the middle POVM elements
makes this POVM a poor candidate indeed for an actual tomographic procedure.
The unit operators mean that the middle outcomes all have nearly the
same probability, with only a weak dependence on the amplitudes $c_j$.
It is from this weak dependence that one must extract the amplitudes,
and this makes a tomographic procedure based on this POVM very
inefficient.

Weigert \cite{Weigert1992a} has described a quite different way of
making a PS\,I-complete measurement, which we can manipulate to produce
a single PS\,I-complete POVM with $2D$ outcomes.  Considering the
system to be a spin-$s$ particle ($D=2s+1$), Weigert first shows that
the probabilities for two spin components, $\vec S\cdot\vec n$ and
$\vec S\cdot\vec n'$, along infinitesimally different directions, are
sufficient to determine a generic pure state up to $2^{2s}=2^{D-1}$
ambiguities.  He then shows that the ambiguities can be resolved by
knowing just the expectation value of the spin component $\vec
S\cdot\vec e$ along the direction $\vec e$ orthogonal to $\vec n$ and
$\vec n'$.

To describe this scheme, let $\ket{\vec n,m}$ denote the eigenvector of
$\vec S\cdot\vec n$ with eigenvalue $m$.  Notice that
\begin{equation}
F\equiv{1\over2s}(\vec S\cdot\vec e+s\mathbb{I})
={1\over2s}\sum_{m=-s}^s(s+m)\ketbra{\vec e,m}{\vec e,m}
\end{equation}
is a potential POVM element whose expectation value contains the same
information as the expectation value of $\vec S\cdot\vec e$.  Weigert
describes getting the expectation value of $\vec S\cdot\vec e$ in the
standard way from measurements of that spin component, but we can get
the same information directly by including $F$ in a POVM. \
Furthermore, since the measurements of $\vec S\cdot\vec n$ and $\vec
S\cdot\vec n'$ are separately normalized, we can amalgamate two
projectors, one from each measurement, into a single POVM element
without losing any information. If we put all this together, Weigert's
work shows that a generic pure state is determined by the expectation
values of the operators
\begin{eqnarray}
&\mbox{}&\ketbra{\vec n,m}{\vec n,m}\;,\quad m=-s+1,\ldots,s,\nonumber\\
&\mbox{}&\ketbra{\vec n',m}{\vec n',m}\;,\quad m=-s+1,\ldots,s,\nonumber\\
&\mbox{}&\ketbra{\vec n,-s}{\vec n,-s}+\ketbra{\vec n',-s}{\vec n',-s}\;,\nonumber\\
&\mbox{}&F\;.
\label{eq:Wlist}
\end{eqnarray}

This list contains $2D$ POVM elements, whose expectation values are
PS\,I-complete, but the list does not constitute a POVM because the
elements don't sum to $\mathbb{I}$.  To turn the list into a POVM, we
use a trick mentioned in the Introduction in the context of IC-POVMs.
Suppose we have a set of positive operators, $\{\Fc\}$, which are
PS\,I-complete in that their expectation values are sufficient to
reconstruct a generic unnormalized vector $\ket{\psi}$ up to a global
phase.  First form the operator $G=\sum_c\Fc$. This operator is
invertible, for if it weren't, there would be a vector $\ket{e}$ such
that $\bra{e}G\ket{e}=0$, which would imply that $\Fc\ket{e}=0$ for all
values of $c\,$; this would mean that the expectation values of the
operators $\Fc$ provide no information about the amplitude
$\braket{e}{\psi}$, so the operators would not be informationally
complete.  Now define new operators $\Ec=G^{-1/2}\Fc G^{-1/2}$.  These
operators are positive and sum to $\mathbb{I}$, so they make up a POVM;
moreover, they are obviously PS\,I-complete.  It is also worth noting
that if $\Fc$ is rank-one, so is $\Ec$.

Although it is easy to calculate $G$ for the list~(\ref{eq:Wlist}), it
is not very enlightening to calculate the POVM elements.  We simply
note that the trick can be applied to the list~(\ref{eq:Wlist}),
thereby turning Weigert's PS\,I-complete measurements into a minimal
PS\,I-complete POVM---and, moreover, one with $2D-2$ rank-one POVM
elements.  The price of this transformation is that the measurement no
longer corresponds to just measuring a few spin components.

One might be tempted to apply this trick to the POVM
elements~(\ref{eq:2DPOVM}), without the throw-away $T$, since the
probability for outcome $T$ never appears in the inversion procedure.
We know this can't work, however, because the result would be a
PS\,I-complete POVM with $2D-1$ elements, contradicting the results of
Sec.~\ref{sec:necessity}.  The reason it doesn't work is that the
inversion procedure assumes normalized pure states, and the trick only
works if the inversion works for normalized and unnormalized states.
For unnormalized states, the throw-away can't be discarded because it
is required to fix the normalization.

\section{Rank-one PS\,I-complete measurement with $3D - 2$ outcomes}
\label{sec:pure}

In this section we consider PS\,I-complete POVMs whose POVM elements are
all of rank one, i.e., multiples of one-dimensional projectors.  We
don't know the minimal number of elements in a rank-one PS\,I-complete
POVM.  We suspect it is close to or even equal to $2D$---the reworked
Weigert measurement discussed in Sec.~\ref{sec:sufficiency} has $2D$
elements, with just two not of rank one---but the best we can do for
the present is $3D-2$ elements, an example of which we present now.

The setting here is the same as in Sec.~\ref{sec:sufficiency}, with
states and operators defined as in Eqs.~(\ref{eq:psi}) and
(\ref{eq:Paulijk}), except that we now allow the states $|\psi\rangle$
to be unnormalized.  Consider the following four states in the subspace
spanned by $\ket{j}$ and $\ket{k}$,
\begin{eqnarray}
|\psi_{jk;0}\rangle&=&|j\rangle\;,\nonumber\\
|\psi_{jk;1}\rangle&=&\cos(\theta/2)|j\rangle+\sin(\theta/2)|k\rangle\;,\nonumber\\
|\psi_{jk;2}\rangle&=&\cos(\theta/2)|j\rangle
+e^{2\pi i/3}\sin(\theta/2)|k\rangle\;,\nonumber\\
|\psi_{jk;3}\rangle&=&\cos(\theta/2)|j\rangle
+e^{-2\pi i/3}\sin(\theta/2)|k\rangle\;,
\label{eq:4states}
\end{eqnarray}
where we assume $0<\theta<\pi$.  Notice that we can write
\begin{equation}
|\psi_{jk;\alpha}\rangle=
\cos(\theta/2)|j\rangle
+e^{2\pi i(\alpha-1)/3}\sin(\theta/2)|k\rangle\;,\quad
\alpha=1,2,3.
\end{equation}
The corresponding one-dimensional projection operators are given by
\begin{eqnarray}
|\psi_{jk;0}\rangle\langle\psi_{jk;0}|&=&
{1\over2}\left(P_{jk}+Z_{jk}\right)\;,\nonumber\\
|\psi_{jk;1}\rangle\langle\psi_{jk;1}|&=&
{1\over2}\left(P_{jk}+X_{jk}\sin\theta+Z_{jk}\cos\theta\right)\;,\nonumber\\
|\psi_{jk;2}\rangle\langle\psi_{jk;2}|&=&
{1\over2}\left(P_{jk}-X_{jk}{1\over2}\sin\theta
+Y_{jk}\sqrt{{3\over2}}\sin\theta+Z_{jk}\cos\theta\right)\;,\nonumber\\
|\psi_{jk;3}\rangle\langle\psi_{jk;3}|&=&
{1\over2}\left(P_{jk}-X_{jk}{1\over2}\sin\theta
-Y_{jk}\sqrt{{3\over2}}\sin\theta+Z_{jk}\cos\theta\right)\;.
\end{eqnarray}
When $\cos\theta=-1/3$, these are the four tetrahedral states in the
two-dimensional subspace spanned by $\ket{j}$ and $\ket{k}$; in this
subspace, they are specified by the Bloch
vectors~(\ref{eq:tetrahedron}).  Notice that
\begin{equation}
\sum_{\alpha=1}^3|\psi_{jk;\alpha}\rangle\langle\psi_{jk;\alpha}|=
{3\over2}(P_{jk}+Z_{jk}\cos\theta)=
3\Bigl(\cos^2(\theta/2)|j\rangle\langle j|+\sin^2(\theta/2)|k\rangle\langle k|\Bigr)\;.
\label{eq:3sum}
\end{equation}

We now consider the following POVM elements, numbering $3D-2$,
\begin{eqnarray}
&\mbox{}&a|0\rangle\langle0|\;,\nonumber\\
&\mbox{}&b|\psi_{j-1,j\,;\,\alpha}\rangle\langle\psi_{j-1,j\,;\,\alpha}|\;,
\quad j=1,\ldots,D-1,\quad \alpha=1,2,3,
\label{eq:ops}
\end{eqnarray}
where we assume $a,b>0$.  The first thing to show is that these POVM
elements are sufficient to reconstruct a generic unnormalized pure
state $\ket{\psi}$.  We begin by determining $c_0=r_0>0$ from the
probability for the first POVM element,
$a|\langle0|\psi\rangle|^2=ar_0^2$.  As we work through the remaining
amplitudes, suppose that we have determined $c_0,\ldots,c_{j-1}$ and
are now trying to get $c_j$ from the three POVM probabilities
\begin{equation}
b\bigl|\langle\psi_{j-1,j\,;\,\alpha}|\psi\rangle\bigr|^2=
b\bigl|\cos(\theta/2)c_{j-1} +e^{-2\pi i(\alpha-1)/3}
\sin(\theta/2)c_j\bigr|^2\;, \quad \alpha=1,2,3.
\label{eq:threeprobs}
\end{equation}
Writing these three probabilities out, it becomes clear that they
determine $c_j$, provided $c_{j-1}\ne0$.  The only hitch in this method
is that we get stuck if any $c_j$ except the last is zero, so we have a
set of $D-1$ $(D-1)$-dimensional subspaces where the method fails. This
failure set is a set of measure zero, however, so the procedure
succeeds in reconstructing a generic pure state.

We now use the trick that we applied to Weigert's measurement in
Sec.~\ref{sec:sufficiency}.  We form the sum of the operators in
Eq.~(\ref{eq:ops}),
\begin{eqnarray}
G&=&a|0\rangle\langle0|+
\sum_{j=1}^{D-1}\sum_{\alpha=1}^3
b|\psi_{j-1,j\,;\,\alpha}\rangle\langle\psi_{j-1,j\,;\,\alpha}|\nonumber\\
&=&\bigl(a+3b\cos^2(\theta/2)\bigr)|0\rangle\langle0|
+3b\sum_{j=1}^{D-2}|j\rangle\langle j|
+3b\sin^2(\theta/2)|D-1\rangle\langle D-1|\;,
\label{eq:G}
\end{eqnarray}
where we use the sum~(\ref{eq:3sum}).  Choosing $b=1/3$ and
$a=\sin^2(\theta/2)$, we get
\begin{equation}
G=\mathbb{I}-\cos^2(\theta/2)|D-1\rangle\langle D-1|\;.
\label{eq:G2}
\end{equation}

The trick introduced in Sec.~\ref{sec:sufficiency} now tells us that
the following POVM elements make up a POVM that is PS\,I-complete:
\begin{eqnarray}
&\mbox{}&\sin^2(\theta/2)G^{-1/2}|0\rangle\langle0|G^{-1/2}
=\sin^2(\theta/2)|0\rangle\langle0|\;,\nonumber\\
&\mbox{}&{1\over3}G^{-1/2}|\psi_{j-1,j\,;\,\alpha}\rangle
\langle\psi_{j-1,j\,;\,\alpha}|G^{-1/2}
={1\over3}|\psi_{j-1,j\,;\,\alpha}\rangle\langle\psi_{j-1,j\,;\,\alpha}|\;,\quad
j=1,\ldots,D-2,\quad\alpha=1,2,3,\nonumber\\
&\mbox{}&{1\over3}G^{-1/2}|\psi_{D-2,D-1;\alpha}\rangle
\langle\psi_{D-2,D-1;\alpha}|G^{-1/2}\;,\quad\alpha=1,2,3.
\label{eq:3D-2}
\end{eqnarray}
Here
\begin{equation}
G^{-1/2}|\psi_{D-2,D-1;\alpha}\rangle =\cos(\theta/2)|D-2\rangle
+e^{2\pi i(\alpha-1)/3}|D-1\rangle\;,\quad\alpha=1,2,3.
\end{equation}
It is easy to see directly, without appealing to the general validity
of our trick, that these particular elements make up a rank-one POVM
that is PS\,I-complete.

Notice that $D=2$ is a special case, because the sum in the second form
of Eq.~(\ref{eq:G}) is missing.  By making the tetrahedral choice of
angle with $a=b=1/2$, we get the four-element tetrahedral POVM of
Eq.~(\ref{eq:tetrahedron}).  Thus the POVM~(\ref{eq:3D-2}) can be
considered as a sort of generalization of the tetrahedral measurement.

Amiet and Weigert \cite{Amiet1999b} have formulated a PS\,I-complete
measurement procedure for a spin-$s$ particle, which is based on
measuring the probabilities for spin components along any three
noncoplanar axes.  By amalgamating three of the outcomes, one from each
of the spin components, this procedure can be reduced to a PS\,I-complete
POVM with $3D-2$ elements, but with the one amalgamated element not of
rank one.

\section{Conclusion}
\label{sec:conclusion}

To determine any pure state of a $D$-dimensional quantum system, except
perhaps a set of measure zero, requires a measurement with at least
$2D$ outcomes.   That is the key result of this paper.  We also
demonstrated that a $3D-2$ element POVM composed only of rank-one
operators (multiples of one-dimensional projectors) is sufficient for
determining a generic pure state.  A number of open questions suggest
themselves.  Is it possible to construct a $2D$-element POVM that can
determine any pure state, without having to exclude sets of measure
zero?  If not, how many extra elements must be added to achieve this
stronger form of PS\,I-completeness?  What is the minimal number of
rank-one POVMs necessary for PS\,I-completeness?  How do the results
obtained here connect with PS\,I-completeness for wave functions of
infinite-dimensional quantum systems?  Finally, we have done next to
nothing in this paper on the question of the efficiency of
PS\,I-complete POVMs for determining pure states from outcome
frequencies in an actual tomographic procedure.  We know that the
$2D$-element PS\,I-complete POVM introduced in
Sec.~\ref{sec:sufficiency} is very inefficient, but we have nothing to
say at present about what efficiencies can be achieved by minimal
PS\,I-complete POVMs.  These questions suggest avenues for research
that might turn PS\,I-complete POVMs into useful tomographic tools in
situations where identical systems are confidently thought to be in a
pure state.

\section{Epilogue}
\label{S:epilogue}

$2D$, or not $2D$: That was the question.  Nature, her pride to
assuage, tempts with another: whether 'tis nobler in the mind to suffer
the slings and arrows of $3D-2$ rank-one POVM elements or to take arms
against a sea of equations, and by opposing, solve them?  Alas, Nature
hath her secrets hidden well; yea, even the mightiest among us hath
failed her veil to penetrate.  Be not discouraged!, dear reader, for we
leave to thee the Herculean task to cast out these $D-2$ elements,
banishing them thus, as so many angels fallen from the Kingdom of
Heaven.  States without States, measuring without Measurements.  {\it
Parvo non ex nihilo\/} \cite{rasputin}. And if, upon completion, $2D$
carries the day: Lo! such beauty to behold! Ne'er shall our POVMs a
finer form take, for it be optimal.  And should $3D-2$ remain, recall:
``A dream itself is but a shadow.''  A shadow of what? ``\dots~a
shadow's shadow."  Thus the Symphony of the Universe plays: movement
upon movement, endlessly subtle, till time itself reigns alone as King.

\acknowledgments

We thank R.~Blume-Kohout for conjecturing that a $2D$-element POVM is
necessary and sufficient for PS\,I-completeness, a conjecture that
started us on this project; D.~Bacon for constructing a PS\,I-complete
POVM with $3D-1$ elements, similar to that in Sec.~\ref{sec:pure}, a
construction that provided the impetus for our work; and D.~Finley for
useful discussions about manifolds.  STF also thanks B.~Eastin and
T.~Johnson for valuable discussions.  This work was supported in part
by US Office of Naval Research Grant No.~N00014-03-1-0426.


\begin{references}

\bibitem{Prugovecki1977a}
E.~Prugove{\v c}ki, Int.\ J.\ Theor.\ Phys.\ {\bf 16}, 321 (1977).

\bibitem{Busch1989a}
P.~Busch and P.~J. Lahti, Found.\ Phys.\ {\bf 19}, 633 (1989).

\bibitem{Peres1993a}
A.~Peres, Quantum Theory: Concepts and Methods (Kluwer Academic,
Dordrecht, The Netherlands, 1993).  POVMs are discussed in Secs.~9-5
and 9-6, and PS\,I-complete measurements in Sec.~3-5.

\bibitem{Caves2002b}
C.~M.~Caves, C.~A. Fuchs, and R.~Schack, J.\ Math.\ Phys.\ {\bf 43},
4537 (2002).

\bibitem{Renes2004a}
J.~M. Renes, R.~Blume-Kohout, A.~J. Scott, and C.~M. Caves, J.\ Math.\
Phys., to be published, {\tt arXiv.org e-print quant-ph/0401149}.

\bibitem{Renes2004b}
J.~M. Renes, ``Spherical code key distribution protocols for qubits,''
{\tt arXiv.org e-print quant-ph/0402135}.

\bibitem{Fuchs2004}
C.~A. Fuchs, ``On the quantumness of a Hilbert space,''
{\tt arXiv.org e-print quant-ph/0404122}.

\bibitem{Fuchs2003}
C.~A. Fuchs and M.~Sasaki, Quant.\ Inf.\ Comp.\ {\bf 3}, 377 (2003).

\bibitem{Pauli}
W.~Pauli, in {\it Handbuch der Physik}, Vol.~XXIV, Pt.~1, edited by
H.~Geiger and K.~Scheel (Springer, Berlin, 1933), p.~98; reprinted in
{\it Encyclopedia of Physics}, Vol.~V, Part 1 (Springer, Berlin, 1958),
p.~17.

\bibitem{Weigert1992a}
S.~Weigert, Phys.\ Rev.~A {\bf 45}, 7688 (1992).

\bibitem{Weigert1996a}
S.~Weigert, Phys.\ Rev.~A {\bf 53}, 2078 (1996).

\bibitem{Corbett1978a}
J.~V. Corbett and C.~A. Hurst, J.\ Austral.\ Math.\ Soc.~B {\bf 20},
182 (1978).

\bibitem{Amiet1999b}
J.-P.~Amiet and S.~Weigert, J.\ Phys.~A {\bf 32}, 2777 (1999).

\bibitem{Amiet1999c}
J.-P.~Amiet and S.~Weigert, J.\ Opt.~B {\bf 1}, L5 (1999).

\bibitem{Peres1992a}
A.~Peres and W.~K. Wootters, Phys.\ Rev.~A {\bf 66}, 1119 (1992).

\bibitem{sard}
S.~Iyanaga and Y.~Kawada, Eds., {\it Encyclopedic Dictionary of Mathematics\/} (Cambridge, MA, MIT Press, 1980), p. 682.

\bibitem{rasputin}
J.~A.~Wyler, Gen. Rel. Grav. {\bf 5}, 175 (1974).

\end{references}
\end{document}